\documentclass[aps,prl,reprint,superscriptaddress,amsmath,amssymb,amsfonts,showpacs]{revtex4-1}

\usepackage{graphicx}
\usepackage{dcolumn}
\usepackage{bm}
\usepackage{booktabs}

\begin{document}

\title{Spontaneous fission modes and lifetimes of super-heavy elements in the nuclear density functional theory}

\author{A. Staszczak}
\affiliation{Institute of Physics, Maria Curie-Sk{\l}odowska University, pl. M. Curie-Sk{\l}odowskiej 1, 20-031 Lublin, Poland}
\affiliation{Department of Physics and Astronomy, University of Tennessee Knoxville, Tennessee 37996, USA}
\affiliation{Oak Ridge National Laboratory, P.O. Box 2008, Oak Ridge, Tennessee 37831, USA}

\author{A. Baran}
\affiliation{Institute of Physics, Maria Curie-Sk{\l}odowska University, pl. M. Curie-Sk{\l}odowskiej 1, 20-031 Lublin, Poland}
\affiliation{Department of Physics and Astronomy, University of Tennessee Knoxville, Tennessee 37996, USA}
\affiliation{Oak Ridge National Laboratory, P.O. Box 2008, Oak Ridge, Tennessee 37831, USA}

\author{W. Nazarewicz}
\affiliation{Department of Physics and Astronomy, University of Tennessee Knoxville, Tennessee 37996, USA}
\affiliation{Oak Ridge National Laboratory, P.O. Box 2008, Oak Ridge, Tennessee 37831, USA}
\affiliation{Theoretical Physics, University of Warsaw, ul. Ho\.za 69, 00-681 Warsaw, Poland}

\date{\today}

\begin{abstract}
Lifetimes of super-heavy (SH)  nuclei are primarily governed by alpha decay and spontaneous fission (SF). Here we study the competing decay modes of even-even SH isotopes  with $108\le Z\le126$ and $148\le N\le188$ using the state-of-the-art self-consistent nuclear density functional theory framework capable of describing the competition between nuclear attraction and electrostatic repulsion. The collective mass tensor  of the fissioning superfluid nucleus is computed by means of  the cranking approximation to the adiabatic time-dependent Hartree-Fock-Bogoliubov approach. Along the path to fission, our calculations allow for the simultaneous breaking of axial and space inversion symmetries; this may result in lowering SF lifetimes by more than seven orders of magnitude in some cases. We predict two competing SF modes: reflection-symmetric and reflection-asymmetric.
The shortest-lived SH isotopes decay by SF; they are expected to lie in a narrow corridor formed by $^{280}$Hs, $^{284}$Fl, and $^{284}_{118}$Uuo that separates the regions of SH nuclei synthesized in ``cold fusion" and ``hot fusion" reactions.
The  region of long-lived SH nuclei is expected to be centered on $^{294}$Ds with a total  half-life  of $\sim$1.5 days.
\end{abstract}


\maketitle

\textit{Introduction}---The SH nuclei represent the limit of nuclear mass and charge; they inhabit the remote corner of the nuclear landscape whose extent is presently unknown. The mere existence of long-lived SH isotopes has been a  fundamental question in science since  Seaborg and Swiatecki coined the notion of the SH ``island of stability" \cite{Sea69}.

Theoretically, it is anticipated that the majority of SH nuclei would fission and/or $\alpha$-decay, but predictions vary from model to model, primarily due to our inability to make accurate predictions of SF half-lives. Here the main uncertainty is our imperfect knowledge of effective nuclear interactions and the strong Coulomb frustration effects due to the interplay between the long-ranged electrostatic repulsion and the short-ranged nuclear force.
By the end of the 1960s, it had been concluded that the existence of the heaviest nuclei with $Z>104$  was primarily determined by the quantum-mechanical shell effects (i.e., single-particle motion of protons and neutrons in quantum orbits) \cite{Mye66,Sob66,*Nil69}. These early microscopic-macroscopic (MM) calculations predicted the nucleus with $Z=114, N=184$ to be the centre of an island of long-lived SH nuclei. This result stayed practically unchallenged until the late 1990s when self-consistent mean-field (SMF)  models, based on realistic effective interactions, were applied to SH nuclei \cite{Cwiok96npaall}. Currently, most theories agree that nuclei around $N=184$ and $Z$ between 114 and 126 should have binding energies strongly lowered by shell effects, forming a region of increased shell stability \cite{Kru00,*Bender01,Cwio05}.

The use of ``hot fusion" reactions with the neutron-rich $^{48}$Ca beam and actinide targets in Dubna resulted in detection of 48 new nuclides with $Z=104-118$ and $A=266-294$ \cite{Oga10, *Oga12, *Oganessian07, *Oganessian12}. Several $\alpha$-decay chains seen in Dubna  were independently verified  \cite{Stav09,*Dul10}. The most significant outcome of these recent  measurements is the observed increase of half-lives with the increasing neutron number -- consistent with the predicted increased stability of SH nuclei when approaching $N=184$. However, the unambiguous identification of the new isotopes still poses a problem because
their $\alpha$-decay chains terminate by SF before reaching the known region of the nuclear chart. The understanding of the competition between $\alpha$-decay and SF channels in SH nuclei is, therefore, of crucial importance for our ability to map the SH region and assess its extent.

The stability  of heavy and SH nuclei is profoundly affected by nuclear deformability through the competing fission valleys having different geometries.  The optimal trajectory in a multidimensional space of collective coordinates  that minimizes the collective action can be associated with  sequences of intrinsic symmetry-breaking transitions. The effects due to breaking of axial symmetry  are known to be important around  the first saddle \cite{Lar72,*Gir83,Baran81all,Moll09all,Sta09all}, and also around the second barrier \cite{Abus12,Lu12}. The reflection-asymmetric mode usually contributes at larger elongations, beyond the first barrier \cite{Nix69,Moll09all,Sta09all,*Baran11a}. The intrinsic symmetry  of the final system -- essential for determining the final split -- depends on the geometry of the post-saddle and pre-scission configurations of the nucleus.

The main objective of this work is to perform realistic predictions of decay modes of SH nuclei using an SMF approach based on the superfluid nuclear density functional theory (DFT) at the deformed Hartree-Fock-Bogoliubov (HFB) level. The advantage of this method is its ability to properly treat the self-consistent interplay between the long-ranged electrostatic repulsion and short-ranged nuclear attraction that gives rise to Coulomb frustration \cite{Cwiok96npaall}. Our calculation -- based on a realistic density dependent effective interaction between nucleons and the microscopic description of the collective action  --  provides a quantitative description of decay properties of known major and minor actinides.  This gives us some confidence in   extrapolations to yet-undiscovered regions of SH nuclei. While several systematic studies of fission barriers of SH nuclei, based on both MM  \cite{Moll09all,Kowa10} and SMF models \cite{Burv04all,Kara10all,Abus12} have been carried out, fission barriers are not observables that can  be directly related to experiment. Moreover,  no microscopic survey of SF properties on SH nuclei exists in the literature, except for some MM studies \cite{Baran81all,Mol89,Stas96all,Smol95,*Smol97,Ghe99} carried out in constrained deformation  spaces and lacking crucial self-consistent polarization effects, and recent  SMF \cite{Erl12,*Schin09,Warda12} calculations  limited by symmetry constraints imposed for most nuclei studied. As we shall point out in this work,  imposing axial and/or space inversion symmetry could result in  overestimation of SF half-lives by many orders of magnitude.

\textit{Model}---The phenomenon of fission can be understood in terms of many-body tunneling involving  mean fields with different intrinsic symmetries \cite{Krappe-Pomorski}.  For SH nuclei, the theoretical tool of choice is the self-consistent nuclear density functional theory (DFT) \cite{Ben03}. The advantage of DFT is that, while treating the nucleus as a many-body system of fermions, it provides a line for identifying the essential collective degrees of freedom and provides a starting point for time-dependent extensions \cite{Gou05}. To describe the quantum-mechanical motion under the collective barrier, it is convenient to employ the adiabatic time-dependent HFB (ATDHFB) theory \cite{Bar78,*Dob81,*Gru82,Ska08} that has been successfully applied to fission \cite{Yul99all,Baran11call}.

The  Skyrme-HFB calculations were carried out using the framework previously discussed in Refs.~\cite{Sta09all,*Sta11,Baran11call}  based on the symmetry unrestricted DFT solver HFODD \cite{hfodd7all} capable of breaking all self-consistent symmetries of nuclear mean fields on the way to fission.   The nuclear energy density functional was approximated by the SkM$^*$ functional \cite{Bart82all} in the particle-hole channel. This functional provides very reasonable results for fission barriers and SF half-lives of even-even actinide nuclei  \cite{Kor12,McDonn12}. In the particle-particle channel, we employed the density-dependent mixed pairing interaction \cite{Dob02}. To truncate  the quasi-particle space of HFB, we adopted the quasiparticle-cut-off value of 60\,MeV in the equivalent energy spectrum \cite{Dob84,*Dob96all}. As discussed in Refs.~\cite{Bor06awall,*Pei11}, such a large value of cut-off energy guarantees the stability of HFB results. The pairing  strengths were adjusted to reproduce the neutron and proton pairing gaps in $^{252}$Fm \cite{Sta09all}; the resulting values are  $V_{n0}=-268.9$\,MeV\,fm$^3$  and $V_{p0}=-332.5$\,MeV\,fm$^3$. The single-particle basis, consisting of the lowest 1140 stretched states originating from the 26 major oscillator shells,fully guarantees the stability of HFODD results~\cite{Sta05a}. All HFB states were taken to compute the mass tensor.

To find the optimum  trajectories in a multidimensional collective space, we  constrain the nuclear collective coordinates associated with  the multipole moments $Q_{\lambda\mu}$, by which we explore the main degrees of freedom related to elongation $(\lambda\mu=20)$, reflection-asymmetry $(\lambda\mu=30)$, triaxiality ($\lambda\mu=22$), and necking $(\lambda\mu=40)$. The driving quadrupole moment $Q_{20}$ is used only as a suitable parameter enumerating consecutive points of the one-dimensional collective path in a multi-dimensional configuration space.

\begin{figure}[htb]
  \includegraphics[width=0.75\columnwidth]{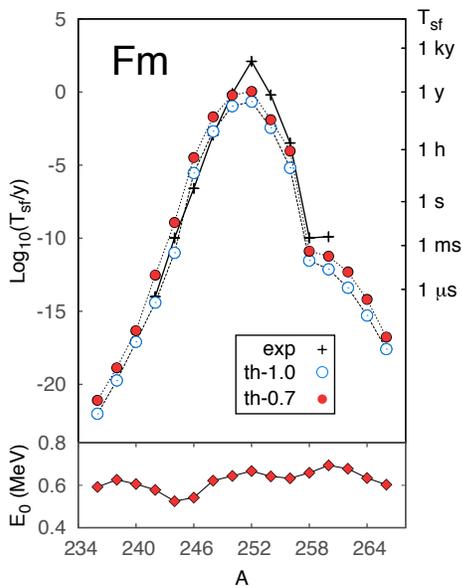}
  \caption{\label{Fig1} (Color online) SF half-lives of even-even Fm isotopes with 236$\leq$A$\leq$266, calculated in this study (th-0.7) compared with experimental data \cite{Hol00,*Khu08}. The corresponding collective ground state energies $E_{0}= 0.7\, \mbox{ZPE}(Q^{gs}_{20})$ are shown in the lower panel. The scaling factor of 0.7  improves the agreement with experimental data. The results obtained without scaling (th-1.0) are also shown.
    }
\end{figure}

The microscopic ingredients needed to compute the action integral and  penetrability are: the collective potential energy, collective inertia (mass tensor),  and collective ground state (g.s.) energy. To calculate the potential energy, we subtract from the total HFB energy $E^{tot}(Q_{20})$ the spurious vibrational zero-point energy ZPE$(Q_{20})$ obtained using the Gaussian overlap approximation as in Refs.~\cite{Staszczak89,*Baran07call}. In this work, we use the perturbative HFB cranking expression for the quadrupole mass parameter $B_{20,20}(Q_{20})$ \cite{Baran11call}. The collective g.s. energy is assumed to be $E_{0}= 0.7\, \mbox{ZPE}(Q^{gs}_{20})$. As shown in Fig.~\ref{Fig1}, the scaling factor of 0.7 improves the agreement between experiment and theory for the SF half-lives of even-even  Fm isotopes. Finally, the  penetrability  has been calculated in WKB according to
Refs.~\cite{Leboeuf73a,*Leboeuf73b,*Leboeuf73c} employing action integrals computed along the static fission pathways.

\textit{Results}---To demonstrate that our model is  capable of describing experimental observations, Fig.~\ref{Fig1} displays predicted SF half-lives for even-even Fm isotopes. This is a challenging case as the measured values \cite{Hol00,*Khu08} vary within this isotopic chain by almost 20 decades. It is satisfying to see a quantitative agreement between experiment and theory. Similar calculations performed for the major and minor actinides \cite{McDonn12} also provide good reproduction of SF half-lives. We wish to stress that a good agreement with existing data is a necessary condition for any model to carry out an extrapolation into the unknown region of SH nuclei.


\begin{figure}[htb]
  \includegraphics[width=\columnwidth]{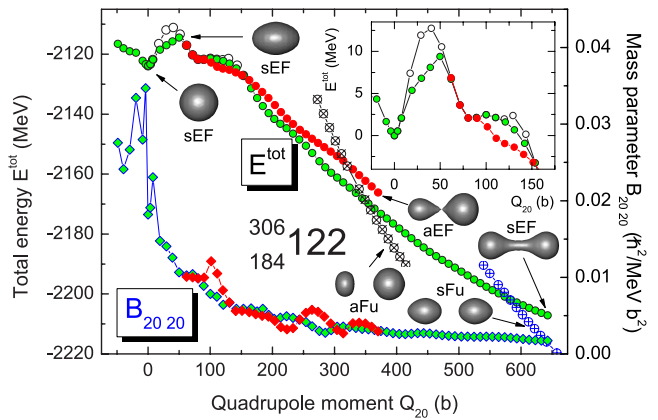}
  \caption{\label{Fig2} (Color online) Total energy  (circles, left scale) and the quadrupole mass parameter (diamonds; right scale) calculated along sEF and aEF fission pathways in $^{306}122$, together with the corresponding shapes. The energy curves along sFu and aFu fusion valleys are also indicated. To illustrate the effect of triaxiality on the inner and outer barrier, the axially symmetric sEF fission pathway is marked by open circles. The deformation energy, normalized with respect to the total  g.s. energy, is shown in the inset.
    }
\end{figure}

The even-even superheavy nuclei with $108\le Z\le126$ and $148\le N\le188$ can be divided into three groups according their g.s. properties \cite{Cwio05,Jach11,Mol12t}: (i) nuclei with prolate-deformed shapes ($Q_{20}\approx30$~b) for $N\le170$; (ii) spherical nuclei for $N>180$; and (iii) weakly deformed, often triaxial systems lying  between (i) and (iii). The  nuclei with $N>180$ are most stable against SF; they have two-humped barriers with the inner saddle at $Q_{20}\approx50$~b  that is higher than the outer one ($E_{A}>E_{B}$). In most cases, triaxiality  substantially reduces $E_A$ \cite{Kowa10,Abus12}. Furthermore, for the reflections-symmetric fission pathways with  elongated fragments  (sEF), triaxiality may also reduce $E_B$ \cite{Abus12,Lu12}.
Typically, the reflection-asymmetric fission valley corresponding to asymmetric elongated fragments (aEF) branches away from the sEF pathway at  $Q_{20}>80$~b beyond the inner saddle. For nuclei with $A>280$ and $Z>108$, the outer barrier vanishes along aEF.
SF half-lives of weakly deformed nuclei from the transitional region (iii) were always calculated relative to the prolate-deformed g.s. Both sEF and aEF fission valleys are taken into account in our calculations. The resulting fission probabilities are combined to give the estimated  SF half-life; the larger penetrability determines the SF mode.

To illustrate  the competition between sEF and aEF fission pathways, Fig.~\ref{Fig2} shows the case of the spherical nucleus $^{306}122$. The energy curves along the reflection symmetric fusion (sFu) and asymmetric fusion (aFu) valleys are also presented. The energy gain due to triaxiality in the region of the first and second saddle can be assessed from the energy curves shown in the inset:   the inner barrier is reduced by  $\sim$3 MeV by triaxiality, and the effect around the second saddle is weaker, around 1 MeV. However, the  outer barrier vanishes altogether along aEF  and this favors the reflection-asymmetric fission mode in $^{306}122$. The total density distributions at pre-scission configurations in aEF and sEF  are shown at $Q_{20}\approx 370$\,b and $Q_{20}\approx 650$\,b, respectively. While the neck rapidly vanishes in aEF,  the symmetric pre-scission region is characterized by an extended neck. Figure~\ref{Fig2} also shows the mass parameters $B_{20,20}$ along sEF and aEF pathways. The two $B_{20,20}$ trajectories  are fairly similar, which indicates that it is the   potential energy (in particular, barrier width and height) that determines the  optimal fission pathway in this case.
The SF half-life along the axially symmetric sEF pathway is $T_{sf} = 10^{13.82}$\,s. Triaxial effects along sEF reduce it to $10^{9.39}$\,s,  and the inclusion of reflection-asymmetric shapes (aEF)  brings the  SF half-life of $^{306}122$ down to $T_{sf} = 10^{6.22}$\,s, which corresponds to an overall reduction of $T_{sf}$ by about seven orders of magnitude.

\begin{figure}[htb]
  \includegraphics[width=0.9\columnwidth]{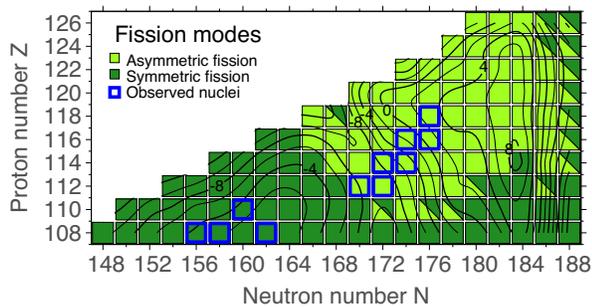}
  \caption{\label{Fig3} (Color online) Competition between sEF and aEF SF modes in even-even SH  nuclei. The bimodal SF is expected in nuclei with $|\log_{10}(T_{\mbox{sEF}}/T_{\mbox{aEF}})|<0.3$ marked by coexisting triangles. The experimentally observed nuclei are indicated. The contours show the predicted SF half-lives in logarithmic scale: $\log_{10}(T_{sf}/s)$.}
 \end{figure}

The survey of the competition between sEF  and aEF SF modes is displayed in Fig.~\ref{Fig3}. The sEF mode dominates for the Hs isotopes, SH nuclei with $A<280$,  and in a triangle defined by  $^{290}$Ds, $^{298}$Fl, and $^{298}$Ds. For the remaining nuclei, the asymmetric  mode is expected to win. In very heavy  nuclei around $N=188$, the bimodal fission is predicted. In Fig.~\ref{Fig3}, the nuclei for which $|\log_{10}(T_{\mbox{sEF}}/T_{\mbox{aEF}})|<0.3$ are marked by triangles. The barrier heights along aEF and sEF are similar; hence, it is  the barrier width that determines the dominant SF mode.

\begin{figure}[htb]
  \includegraphics[width=0.9\columnwidth]{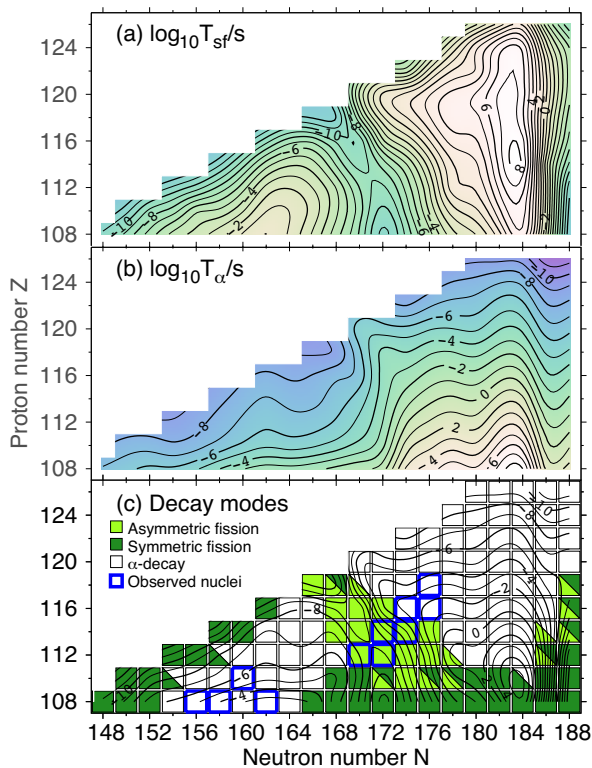}
  \caption{\label{Fig4} (Color online) Summary of our SkM$^*$ results for decay modes of SH nuclei. (a)  SF half-lives $\log_{10}T_{sf}$ (in seconds). (b) $\alpha$-decay  half-lives $\log_{10}T_{\alpha}$ (in seconds). (c) Dominant decay modes. If two modes compete, this is marked by coexisting triangles.}
 \end{figure}
Figure~\ref{Fig4} summarizes our predictions for SF and alpha decay modes of SH nuclei. The  calculated SF half-lives are shown in Fig.~\ref{Fig4}(a). The maximum value of $T_{sf}=10^{7.76}$\,s corresponds to $^{298}$Fl and $T_{sf}$ values of $^{300}$Lv, and $^{302}_{120}$Udn are similar. The shortest SF half-lives, reaching down to 10$^{-10}$\,s, are predicted for nuclei from a narrow corridor formed by $^{280}$Hs, $^{284}$Fl, and $^{284}_{118}$Uuo that lies on the border of weakly-deformed SH nuclei that exhibit prolate-oblate coexistence effects, or g.s. triaxiality \cite{Cwio05,Jach11,Mol12t}. This corridor of fission instability separates the regions of SH nuclei created in hot- and cold-fusion reactions.

It is instructive to compare our SF half-lives  with other predictions. The  MM calculations \cite{Smol95,*Smol97}  yield  SF half-lives that overshoot our results by  more then five orders of magnitude. We attribute this to the assumption of axiality and reflection-symmetry used in their work. Likewise, the axially symmetric HFB+D1S calculation \cite{Warda12} overestimates our SF half-lives by many  orders of magnitude. On the other hand, there is a nice consistency between our aEF results and those
obtained in the axial Skyrme HF+BCS approach of Refs.~\cite{Erl12,*Schin09} with SV-min and SV-bas functionals. In particular, the corridor of the maximum SF instability is predicted similarly by both approaches. It is anticipated, however, that the inclusion of triaxiality is likely to reduce their SF half-lives significantly.

To estimate $\alpha$-decay half-lives, we used the standard Viola-Seaborg expression \cite{Viola66} with the parameters from Ref.~\cite{Sobi05} and calculated $Q_\alpha$ values. Figure~\ref{Fig4}(b) summarizes our results. In general, we  obtain a fairly good agreement with experiment. Our model underestimates experimental $Q_\alpha$ values  in the vicinity of the deformed  shell closure $N=162$. In this respect, the models of Refs. \cite{Smol97,Warda12} are closer to the data.

Our survey of lifetimes of even-even SH nuclei is summarized  in Fig.~\ref{Fig4}(c). According to our model, the region of long-lived SH nuclei is expected to be centered on $^{294}$Ds with a predicted total half-life (considering SF and alpha decay) of $10^{5.13}$ s, i.e.,$\sim$1.5 days. For comparison, the total half-life $^{292}$Ds predicted in Refs. \cite{Smol97,Warda12}  is 51~y and 14 y,
respectively.

In summary, we carried out self-consistent Skyrme-HFB calculations to predict main decay modes of even-even SH nuclei with  $108\le Z\le126$ and $148\le  N\le188$, assess their lifetimes, and estimate the center of enhanced stability in the SH region. In our model, fission pathways in the collective space are not constrained by imposed self-consistent symmetries, and ATDHFB cranking fission inertia and zero-point energy corrections have been obtained microscopically. The model satisfactorily reproduces experimental $T_{sf}$ values in even-even  actinides, which is a necessary condition for a model-based extrapolation to unknown SH nuclei. We wish to emphasize that our survey is the first systematic self-consistent approach to SF in SH nuclei that is free from artificially-imposed symmetry constraints that are likely to affect previous predictions.

We predict two competing SF modes in SH nuclei: the reflection-symmetric mode sEF and the reflection-asymmetric mode aEF. The latter one is expected to prevail  for  $N\ge166$, while sEF shows up in the  region of light SH nuclei and also for neutron-rich nuclei with  $N\approx 188$. The region of asymmetric fission roughly corresponds  to the  region of the highest SF barriers and longest SF half-lives. The predicted SF half-lives of even-even transitional nuclei around $^{284}$Cn are dramatically reduced as compared to the current experimental estimates \cite{Oga10}. Since those systems belong to the region of shape coexistence and/or oblate g.s. shapes, some further increase of SF half-lives is anticipated due to the lowering of g.s. energy due to the shape mixing and/or appearance of a triaxial saddle at low-deformations \cite{Ghe99,Cwio05,Mol12t}. Other  improvements  of the current model include dynamical treatment of  penetrability by considering   several collective coordinates, improved energy density functionals \cite{Kor12}, and the full ATHFB inertia  \cite{Baran11call}.  Work along these lines is in progress.

\begin{acknowledgments}
This work was supported  by the National Nuclear Security Administration under the Stewardship Science Academic Alliances program through DOE Grant DE-FG52-09NA29461; by the U.S. Department of Energy under Contract Nos. DE-FG02-96ER40963 (University of Tennessee); by the NEUP grant DE-AC07-05ID14517 (sub award 00091100); and by the National Science Center (Poland) under Contract
DEC-2011/01/B/ST2/03667.
\end{acknowledgments}

%

\begin{thebibliography}{65}%
\makeatletter
\providecommand \@ifxundefined [1]{%
 \@ifx{#1\undefined}
}%
\providecommand \@ifnum [1]{%
 \ifnum #1\expandafter \@firstoftwo
 \else \expandafter \@secondoftwo
 \fi
}%
\providecommand \@ifx [1]{%
 \ifx #1\expandafter \@firstoftwo
 \else \expandafter \@secondoftwo
 \fi
}%
\providecommand \natexlab [1]{#1}%
\providecommand \enquote  [1]{``#1''}%
\providecommand \bibnamefont  [1]{#1}%
\providecommand \bibfnamefont [1]{#1}%
\providecommand \citenamefont [1]{#1}%
\providecommand \href@noop [0]{\@secondoftwo}%
\providecommand \href [0]{\begingroup \@sanitize@url \@href}%
\providecommand \@href[1]{\@@startlink{#1}\@@href}%
\providecommand \@@href[1]{\endgroup#1\@@endlink}%
\providecommand \@sanitize@url [0]{\catcode `\\12\catcode `\$12\catcode
  `\&12\catcode `\#12\catcode `\^12\catcode `\_12\catcode `\%12\relax}%
\providecommand \@@startlink[1]{}%
\providecommand \@@endlink[0]{}%
\providecommand \url  [0]{\begingroup\@sanitize@url \@url }%
\providecommand \@url [1]{\endgroup\@href {#1}{\urlprefix }}%
\providecommand \urlprefix  [0]{URL }%
\providecommand \Eprint [0]{\href }%
\providecommand \doibase [0]{http://dx.doi.org/}%
\providecommand \selectlanguage [0]{\@gobble}%
\providecommand \bibinfo  [0]{\@secondoftwo}%
\providecommand \bibfield  [0]{\@secondoftwo}%
\providecommand \translation [1]{[#1]}%
\providecommand \BibitemOpen [0]{}%
\providecommand \bibitemStop [0]{}%
\providecommand \bibitemNoStop [0]{.\EOS\space}%
\providecommand \EOS [0]{\spacefactor3000\relax}%
\providecommand \BibitemShut  [1]{\csname bibitem#1\endcsname}%
\let\auto@bib@innerbib\@empty
\bibitem [{\citenamefont {Seaborg}(1969)}]{Sea69}%
  \BibitemOpen
  \bibfield  {author} {\bibinfo {author} {\bibfnamefont {G.~T.}\ \bibnamefont
  {Seaborg}},\ }\href@noop {} {\bibfield  {journal} {\bibinfo  {journal} {J.
  Chem. Educ.}\ }\textbf {\bibinfo {volume} {46}},\ \bibinfo {pages} {626}
  (\bibinfo {year} {1969})}\BibitemShut {NoStop}%
\bibitem [{\citenamefont {Myers}\ and\ \citenamefont
  {Swiatecki}(1966)}]{Mye66}%
  \BibitemOpen
  \bibfield  {author} {\bibinfo {author} {\bibfnamefont {W.~D.}\ \bibnamefont
  {Myers}}\ and\ \bibinfo {author} {\bibfnamefont {W.~J.}\ \bibnamefont
  {Swiatecki}},\ }\href@noop {} {\bibfield  {journal} {\bibinfo  {journal}
  {Nucl. Phys.}\ }\textbf {\bibinfo {volume} {81}},\ \bibinfo {pages} {1}
  (\bibinfo {year} {1966})}\BibitemShut {NoStop}%
\bibitem [{\citenamefont {Sobiczewski}\ \emph {et~al.}(1966)\citenamefont
  {Sobiczewski}, \citenamefont {Gareev},\ and\ \citenamefont
  {Kalinkin}}]{Sob66}%
  \BibitemOpen
  \bibfield  {author} {\bibinfo {author} {\bibfnamefont {A.}~\bibnamefont
  {Sobiczewski}}, \bibinfo {author} {\bibfnamefont {F.~A.}\ \bibnamefont
  {Gareev}}, \ and\ \bibinfo {author} {\bibfnamefont {B.~N.}\ \bibnamefont
  {Kalinkin}},\ }\href@noop {} {\bibfield  {journal} {\bibinfo  {journal}
  {Phys. Lett.}\ }\textbf {\bibinfo {volume} {22}},\ \bibinfo {pages} {500}
  (\bibinfo {year} {1966})}\BibitemShut {NoStop}%
\bibitem [{\citenamefont {Nilsson}\ \emph {et~al.}(1969)\citenamefont {Nilsson}
  \emph {et~al.}}]{Nil69}%
  \BibitemOpen
  \bibfield  {author} {\bibinfo {author} {\bibfnamefont {S.}~\bibnamefont
  {Nilsson}} \emph {et~al.},\ }\href@noop {} {\bibfield  {journal} {\bibinfo
  {journal} {Nucl. Phys. A}\ }\textbf {\bibinfo {volume} {131}},\ \bibinfo
  {pages} {1} (\bibinfo {year} {1969})}\BibitemShut {NoStop}%
\bibitem [{\citenamefont {{\'C}wiok}\ \emph {et~al.}(1996)\citenamefont
  {{\'C}wiok}, \citenamefont {Dobaczewski}, \citenamefont {Heenen},
  \citenamefont {Magierski},\ and\ \citenamefont {Nazarewicz}}]{Cwiok96npaall}%
  \BibitemOpen
  \bibfield  {author} {\bibinfo {author} {\bibfnamefont {S.}~\bibnamefont
  {{\'C}wiok}}, \bibinfo {author} {\bibfnamefont {J.}~\bibnamefont
  {Dobaczewski}}, \bibinfo {author} {\bibfnamefont {P.-H.}\ \bibnamefont
  {Heenen}}, \bibinfo {author} {\bibfnamefont {P.}~\bibnamefont {Magierski}}, \
  and\ \bibinfo {author} {\bibfnamefont {W.}~\bibnamefont {Nazarewicz}},\
  }\href@noop {} {\bibfield  {journal} {\bibinfo  {journal} {Nucl. Phys. A}\
  }\textbf {\bibinfo {volume} {611}},\ \bibinfo {pages} {211} (\bibinfo {year}
  {1996})}\BibitemShut {NoStop}%
\bibitem [{\citenamefont {Kruppa}\ \emph {et~al.}(2000)\citenamefont {Kruppa},
  \citenamefont {Bender}, \citenamefont {Nazarewicz}, \citenamefont {Reinhard},
  \citenamefont {Vertse},\ and\ \citenamefont {\ifmmode~\acute{C}\else
  \'{C}\fi{}wiok}}]{Kru00}%
  \BibitemOpen
  \bibfield  {author} {\bibinfo {author} {\bibfnamefont {A.~T.}\ \bibnamefont
  {Kruppa}}, \bibinfo {author} {\bibfnamefont {M.}~\bibnamefont {Bender}},
  \bibinfo {author} {\bibfnamefont {W.}~\bibnamefont {Nazarewicz}}, \bibinfo
  {author} {\bibfnamefont {P.-G.}\ \bibnamefont {Reinhard}}, \bibinfo {author}
  {\bibfnamefont {T.}~\bibnamefont {Vertse}}, \ and\ \bibinfo {author}
  {\bibfnamefont {S.}~\bibnamefont {\ifmmode~\acute{C}\else \'{C}\fi{}wiok}},\
  }\href@noop {} {\bibfield  {journal} {\bibinfo  {journal} {Phys. Rev. C}\
  }\textbf {\bibinfo {volume} {61}},\ \bibinfo {pages} {034313} (\bibinfo
  {year} {2000})}\BibitemShut {NoStop}%
\bibitem [{\citenamefont {Bender}\ \emph {et~al.}(2001)\citenamefont {Bender},
  \citenamefont {Nazarewicz},\ and\ \citenamefont {Reinhard}}]{Bender01}%
  \BibitemOpen
  \bibfield  {author} {\bibinfo {author} {\bibfnamefont {M.}~\bibnamefont
  {Bender}}, \bibinfo {author} {\bibfnamefont {W.}~\bibnamefont {Nazarewicz}},
  \ and\ \bibinfo {author} {\bibfnamefont {P.-G.}\ \bibnamefont {Reinhard}},\
  }\href@noop {} {\bibfield  {journal} {\bibinfo  {journal} {Phys. Lett. B}\
  }\textbf {\bibinfo {volume} {515}},\ \bibinfo {pages} {42} (\bibinfo {year}
  {2001})}\BibitemShut {NoStop}%
\bibitem [{\citenamefont {{\'C}wiok}\ \emph {et~al.}(2005)\citenamefont
  {{\'C}wiok}, \citenamefont {Heenen},\ and\ \citenamefont
  {Nazarewicz}}]{Cwio05}%
  \BibitemOpen
  \bibfield  {author} {\bibinfo {author} {\bibfnamefont {S.}~\bibnamefont
  {{\'C}wiok}}, \bibinfo {author} {\bibfnamefont {P.-H.}\ \bibnamefont
  {Heenen}}, \ and\ \bibinfo {author} {\bibfnamefont {W.}~\bibnamefont
  {Nazarewicz}},\ }\href@noop {} {\bibfield  {journal} {\bibinfo  {journal}
  {Nature}\ }\textbf {\bibinfo {volume} {433}},\ \bibinfo {pages} {705}
  (\bibinfo {year} {2005})}\BibitemShut {NoStop}%
\bibitem [{\citenamefont {Oganessian}\ \emph {et~al.}(2010)\citenamefont
  {Oganessian} \emph {et~al.}}]{Oga10}%
  \BibitemOpen
  \bibfield  {author} {\bibinfo {author} {\bibfnamefont {Y.~T.}\ \bibnamefont
  {Oganessian}} \emph {et~al.},\ }\href@noop {} {\bibfield  {journal} {\bibinfo
   {journal} {Phys. Rev. Lett.}\ }\textbf {\bibinfo {volume} {104}},\ \bibinfo
  {pages} {142502} (\bibinfo {year} {2010})}\BibitemShut {NoStop}%
\bibitem [{\citenamefont {Oganessian}\ \emph {et~al.}(2012)\citenamefont
  {Oganessian} \emph {et~al.}}]{Oga12}%
  \BibitemOpen
  \bibfield  {author} {\bibinfo {author} {\bibfnamefont {Y.~T.}\ \bibnamefont
  {Oganessian}} \emph {et~al.},\ }\href@noop {} {\bibfield  {journal} {\bibinfo
   {journal} {Phys. Rev. Lett.}\ }\textbf {\bibinfo {volume} {108}},\ \bibinfo
  {pages} {022502} (\bibinfo {year} {2012})}\BibitemShut {NoStop}%
\bibitem [{\citenamefont {Oganessian}(2007)}]{Oganessian07}%
  \BibitemOpen
  \bibfield  {author} {\bibinfo {author} {\bibfnamefont {Y.~T.}\ \bibnamefont
  {Oganessian}},\ }\href@noop {} {\bibfield  {journal} {\bibinfo  {journal} {J.
  Phys. G: Nucl. Part. Phys.}\ }\textbf {\bibinfo {volume} {34}},\ \bibinfo
  {pages} {R165} (\bibinfo {year} {2007})}\BibitemShut {NoStop}%
\bibitem [{\citenamefont {Oganessian}(2012)}]{Oganessian12}%
  \BibitemOpen
  \bibfield  {author} {\bibinfo {author} {\bibfnamefont {Y.~T.}\ \bibnamefont
  {Oganessian}},\ }\href@noop {} {\bibfield  {journal} {\bibinfo  {journal}
  {Acta Phys. Pol. B}\ }\textbf {\bibinfo {volume} {43}},\ \bibinfo {pages}
  {167} (\bibinfo {year} {2012})}\BibitemShut {NoStop}%
\bibitem [{\citenamefont {Stavsetra}\ \emph {et~al.}(2009)\citenamefont
  {Stavsetra} \emph {et~al.}}]{Stav09}%
  \BibitemOpen
  \bibfield  {author} {\bibinfo {author} {\bibfnamefont {L.}~\bibnamefont
  {Stavsetra}} \emph {et~al.},\ }\href@noop {} {\bibfield  {journal} {\bibinfo
  {journal} {Phys. Rev. Lett.}\ }\textbf {\bibinfo {volume} {103}},\ \bibinfo
  {pages} {132502} (\bibinfo {year} {2009})}\BibitemShut {NoStop}%
\bibitem [{\citenamefont {D{\"u}llmann}\ \emph {et~al.}(2010)\citenamefont
  {D{\"u}llmann} \emph {et~al.}}]{Dul10}%
  \BibitemOpen
  \bibfield  {author} {\bibinfo {author} {\bibfnamefont {C.~E.}\ \bibnamefont
  {D{\"u}llmann}} \emph {et~al.},\ }\href@noop {} {\bibfield  {journal}
  {\bibinfo  {journal} {Phys. Rev. Lett.}\ }\textbf {\bibinfo {volume} {104}},\
  \bibinfo {pages} {252701} (\bibinfo {year} {2010})}\BibitemShut {NoStop}%
\bibitem [{\citenamefont {Larsson}\ \emph {et~al.}(1972)\citenamefont
  {Larsson}, \citenamefont {Ragnarsson},\ and\ \citenamefont
  {Nilsson}}]{Lar72}%
  \BibitemOpen
  \bibfield  {author} {\bibinfo {author} {\bibfnamefont {S.~E.}\ \bibnamefont
  {Larsson}}, \bibinfo {author} {\bibfnamefont {I.}~\bibnamefont {Ragnarsson}},
  \ and\ \bibinfo {author} {\bibfnamefont {S.~G.}\ \bibnamefont {Nilsson}},\
  }\href@noop {} {\bibfield  {journal} {\bibinfo  {journal} {Phys. Lett. B}\
  }\textbf {\bibinfo {volume} {38}},\ \bibinfo {pages} {269} (\bibinfo {year}
  {1972})}\BibitemShut {NoStop}%
\bibitem [{\citenamefont {Girod}\ and\ \citenamefont
  {Grammaticos}(1983)}]{Gir83}%
  \BibitemOpen
  \bibfield  {author} {\bibinfo {author} {\bibfnamefont {M.}~\bibnamefont
  {Girod}}\ and\ \bibinfo {author} {\bibfnamefont {B.}~\bibnamefont
  {Grammaticos}},\ }\href@noop {} {\bibfield  {journal} {\bibinfo  {journal}
  {Phys. Rev. C}\ }\textbf {\bibinfo {volume} {27}},\ \bibinfo {pages} {2317}
  (\bibinfo {year} {1983})}\BibitemShut {NoStop}%
\bibitem [{\citenamefont {Baran}\ \emph {et~al.}(1981)\citenamefont {Baran},
  \citenamefont {Pomorski}, \citenamefont {{\L}ukasiak},\ and\ \citenamefont
  {Sobiczewski}}]{Baran81all}%
  \BibitemOpen
  \bibfield  {author} {\bibinfo {author} {\bibfnamefont {A.}~\bibnamefont
  {Baran}}, \bibinfo {author} {\bibfnamefont {K.}~\bibnamefont {Pomorski}},
  \bibinfo {author} {\bibfnamefont {A.}~\bibnamefont {{\L}ukasiak}}, \ and\
  \bibinfo {author} {\bibfnamefont {A.}~\bibnamefont {Sobiczewski}},\
  }\href@noop {} {\bibfield  {journal} {\bibinfo  {journal} {Nucl. Phys. A}\
  }\textbf {\bibinfo {volume} {361}},\ \bibinfo {pages} {83} (\bibinfo {year}
  {1981})}\BibitemShut {NoStop}%
\bibitem [{\citenamefont {M\"oller}\ \emph {et~al.}(2009)\citenamefont
  {M\"oller}, \citenamefont {Sierk}, \citenamefont {Ichikawa}, \citenamefont
  {Iwamoto}, \citenamefont {Bengtsson}, \citenamefont {Uhrenholt},\ and\
  \citenamefont {{{\AA}}berg}}]{Moll09all}%
  \BibitemOpen
  \bibfield  {author} {\bibinfo {author} {\bibfnamefont {P.}~\bibnamefont
  {M\"oller}}, \bibinfo {author} {\bibfnamefont {A.~J.}\ \bibnamefont {Sierk}},
  \bibinfo {author} {\bibfnamefont {T.}~\bibnamefont {Ichikawa}}, \bibinfo
  {author} {\bibfnamefont {A.}~\bibnamefont {Iwamoto}}, \bibinfo {author}
  {\bibfnamefont {R.}~\bibnamefont {Bengtsson}}, \bibinfo {author}
  {\bibfnamefont {H.}~\bibnamefont {Uhrenholt}}, \ and\ \bibinfo {author}
  {\bibfnamefont {S.}~\bibnamefont {{{\AA}}berg}},\ }\href@noop {} {\bibfield
  {journal} {\bibinfo  {journal} {Phys. Rev. C}\ }\textbf {\bibinfo {volume}
  {79}},\ \bibinfo {pages} {064304} (\bibinfo {year} {2009})}\BibitemShut
  {NoStop}%
\bibitem [{\citenamefont {Staszczak}\ \emph {et~al.}(2009)\citenamefont
  {Staszczak}, \citenamefont {Baran}, \citenamefont {Dobaczewski},\ and\
  \citenamefont {Nazarewicz}}]{Sta09all}%
  \BibitemOpen
  \bibfield  {author} {\bibinfo {author} {\bibfnamefont {A.}~\bibnamefont
  {Staszczak}}, \bibinfo {author} {\bibfnamefont {A.}~\bibnamefont {Baran}},
  \bibinfo {author} {\bibfnamefont {J.}~\bibnamefont {Dobaczewski}}, \ and\
  \bibinfo {author} {\bibfnamefont {W.}~\bibnamefont {Nazarewicz}},\
  }\href@noop {} {\bibfield  {journal} {\bibinfo  {journal} {Phys.\ Rev. C}\
  }\textbf {\bibinfo {volume} {80}},\ \bibinfo {pages} {014309} (\bibinfo
  {year} {2009})}\BibitemShut {NoStop}%
\bibitem [{\citenamefont {Abusara}\ \emph {et~al.}(2012)\citenamefont
  {Abusara}, \citenamefont {Afanasjev},\ and\ \citenamefont {Ring}}]{Abus12}%
  \BibitemOpen
  \bibfield  {author} {\bibinfo {author} {\bibfnamefont {H.}~\bibnamefont
  {Abusara}}, \bibinfo {author} {\bibfnamefont {A.}~\bibnamefont {Afanasjev}},
  \ and\ \bibinfo {author} {\bibfnamefont {P.}~\bibnamefont {Ring}},\
  }\href@noop {} {\bibfield  {journal} {\bibinfo  {journal} {Phys. Rev. C}\
  }\textbf {\bibinfo {volume} {85}},\ \bibinfo {pages} {024314} (\bibinfo
  {year} {2012})}\BibitemShut {NoStop}%
\bibitem [{\citenamefont {Lu}\ \emph {et~al.}(2012)\citenamefont {Lu},
  \citenamefont {Zhao},\ and\ \citenamefont {Zhou}}]{Lu12}%
  \BibitemOpen
  \bibfield  {author} {\bibinfo {author} {\bibfnamefont {B.-N.}\ \bibnamefont
  {Lu}}, \bibinfo {author} {\bibfnamefont {E.-G.}\ \bibnamefont {Zhao}}, \ and\
  \bibinfo {author} {\bibfnamefont {S.-G.}\ \bibnamefont {Zhou}},\ }\href@noop
  {} {\bibfield  {journal} {\bibinfo  {journal} {Phys. Rev. C}\ }\textbf
  {\bibinfo {volume} {85}},\ \bibinfo {pages} {011301(R)} (\bibinfo {year}
  {2012})}\BibitemShut {NoStop}%
\bibitem [{\citenamefont {Nix}(1969)}]{Nix69}%
  \BibitemOpen
  \bibfield  {author} {\bibinfo {author} {\bibfnamefont {J.~R.}\ \bibnamefont
  {Nix}},\ }\href@noop {} {\bibfield  {journal} {\bibinfo  {journal} {Nucl.
  Phys. A}\ }\textbf {\bibinfo {volume} {130}},\ \bibinfo {pages} {241}
  (\bibinfo {year} {1969})}\BibitemShut {NoStop}%
\bibitem [{\citenamefont {Staszczak}\ \emph
  {et~al.}(2011{\natexlab{a}})\citenamefont {Staszczak}, \citenamefont
  {Baran},\ and\ \citenamefont {Nazarewicz}}]{Baran11a}%
  \BibitemOpen
  \bibfield  {author} {\bibinfo {author} {\bibfnamefont {A.}~\bibnamefont
  {Staszczak}}, \bibinfo {author} {\bibfnamefont {A.}~\bibnamefont {Baran}}, \
  and\ \bibinfo {author} {\bibfnamefont {W.}~\bibnamefont {Nazarewicz}},\
  }\href@noop {} {\bibfield  {journal} {\bibinfo  {journal} {Int. J. Mod. Phys.
  E}\ }\textbf {\bibinfo {volume} {20}},\ \bibinfo {pages} {552} (\bibinfo
  {year} {2011}{\natexlab{a}})}\BibitemShut {NoStop}%
\bibitem [{\citenamefont {Kowal}\ \emph {et~al.}(2010)\citenamefont {Kowal},
  \citenamefont {Jachimowicz},\ and\ \citenamefont {Sobiczewski}}]{Kowa10}%
  \BibitemOpen
  \bibfield  {author} {\bibinfo {author} {\bibfnamefont {M.}~\bibnamefont
  {Kowal}}, \bibinfo {author} {\bibfnamefont {P.}~\bibnamefont {Jachimowicz}},
  \ and\ \bibinfo {author} {\bibfnamefont {A.}~\bibnamefont {Sobiczewski}},\
  }\href@noop {} {\bibfield  {journal} {\bibinfo  {journal} {Phys. Rev. C}\
  }\textbf {\bibinfo {volume} {82}},\ \bibinfo {pages} {014303} (\bibinfo
  {year} {2010})}\BibitemShut {NoStop}%
\bibitem [{\citenamefont {B\"urvenich}\ \emph {et~al.}(2004)\citenamefont
  {B\"urvenich}, \citenamefont {Bender}, \citenamefont {Maruhn},\ and\
  \citenamefont {Reinhard}}]{Burv04all}%
  \BibitemOpen
  \bibfield  {author} {\bibinfo {author} {\bibfnamefont {T.}~\bibnamefont
  {B\"urvenich}}, \bibinfo {author} {\bibfnamefont {M.}~\bibnamefont {Bender}},
  \bibinfo {author} {\bibfnamefont {J.~A.}\ \bibnamefont {Maruhn}}, \ and\
  \bibinfo {author} {\bibfnamefont {P.-G.}\ \bibnamefont {Reinhard}},\
  }\href@noop {} {\bibfield  {journal} {\bibinfo  {journal} {Phys. Rev. C}\
  }\textbf {\bibinfo {volume} {69}},\ \bibinfo {pages} {014307} (\bibinfo
  {year} {2004})}\BibitemShut {NoStop}%
\bibitem [{\citenamefont {Karatzikos}\ \emph {et~al.}(2010)\citenamefont
  {Karatzikos}, \citenamefont {Afanasjev}, \citenamefont {Lalazissis},\ and\
  \citenamefont {Ring}}]{Kara10all}%
  \BibitemOpen
  \bibfield  {author} {\bibinfo {author} {\bibfnamefont {S.}~\bibnamefont
  {Karatzikos}}, \bibinfo {author} {\bibfnamefont {A.}~\bibnamefont
  {Afanasjev}}, \bibinfo {author} {\bibfnamefont {G.}~\bibnamefont
  {Lalazissis}}, \ and\ \bibinfo {author} {\bibfnamefont {P.}~\bibnamefont
  {Ring}},\ }\href@noop {} {\bibfield  {journal} {\bibinfo  {journal} {Phys.
  Lett. B}\ }\textbf {\bibinfo {volume} {689}},\ \bibinfo {pages} {72}
  (\bibinfo {year} {2010})}\BibitemShut {NoStop}%
\bibitem [{\citenamefont {M{\"o}ller}\ \emph {et~al.}(1989)\citenamefont
  {M{\"o}ller}, \citenamefont {Nix},\ and\ \citenamefont {Swiatecki}}]{Mol89}%
  \BibitemOpen
  \bibfield  {author} {\bibinfo {author} {\bibfnamefont {P.}~\bibnamefont
  {M{\"o}ller}}, \bibinfo {author} {\bibfnamefont {J.}~\bibnamefont {Nix}}, \
  and\ \bibinfo {author} {\bibfnamefont {W.}~\bibnamefont {Swiatecki}},\
  }\href@noop {} {\bibfield  {journal} {\bibinfo  {journal} {Nucl. Phys. A}\
  }\textbf {\bibinfo {volume} {492}},\ \bibinfo {pages} {349} (\bibinfo {year}
  {1989})}\BibitemShut {NoStop}%
\bibitem [{\citenamefont {Staszczak}\ \emph {et~al.}(1996)\citenamefont
  {Staszczak}, \citenamefont {{\L}ojewski}, \citenamefont {Baran},
  \citenamefont {Nerlo-Pomorska},\ and\ \citenamefont {Pomorski}}]{Stas96all}%
  \BibitemOpen
  \bibfield  {author} {\bibinfo {author} {\bibfnamefont {A.}~\bibnamefont
  {Staszczak}}, \bibinfo {author} {\bibfnamefont {Z.}~\bibnamefont
  {{\L}ojewski}}, \bibinfo {author} {\bibfnamefont {A.}~\bibnamefont {Baran}},
  \bibinfo {author} {\bibfnamefont {B.}~\bibnamefont {Nerlo-Pomorska}}, \ and\
  \bibinfo {author} {\bibfnamefont {K.}~\bibnamefont {Pomorski}},\ }in\
  \href@noop {} {\emph {\bibinfo {booktitle} {{Proc. 3rd Int. Conf.,
  \v{C}ast{\'a}-Papierni\v{c}ka, 1996}}}},\ Vol.\ \bibinfo {volume} {Dynamical
  Aspects of Nuclear Fission},\ \bibinfo {editor} {edited by\ \bibinfo {editor}
  {\bibfnamefont {J.}~\bibnamefont {Kliman}}\ and\ \bibinfo {editor}
  {\bibfnamefont {B.~I.}\ \bibnamefont {Pustylnik}}}\ (\bibinfo  {publisher}
  {JINR, Dubna},\ \bibinfo {year} {1996})\ p.~\bibinfo {pages} {22}\BibitemShut
  {NoStop}%
\bibitem [{\citenamefont {Smola{\'n}czuk}\ \emph {et~al.}(1995)\citenamefont
  {Smola{\'n}czuk}, \citenamefont {Skalski},\ and\ \citenamefont
  {Sobiczewski}}]{Smol95}%
  \BibitemOpen
  \bibfield  {author} {\bibinfo {author} {\bibfnamefont {R.}~\bibnamefont
  {Smola{\'n}czuk}}, \bibinfo {author} {\bibfnamefont {J.}~\bibnamefont
  {Skalski}}, \ and\ \bibinfo {author} {\bibfnamefont {A.}~\bibnamefont
  {Sobiczewski}},\ }\href@noop {} {\bibfield  {journal} {\bibinfo  {journal}
  {Phys. Rev. C}\ }\textbf {\bibinfo {volume} {52}},\ \bibinfo {pages} {1871}
  (\bibinfo {year} {1995})}\BibitemShut {NoStop}%
\bibitem [{\citenamefont {Smola{\'n}czuk}(1997)}]{Smol97}%
  \BibitemOpen
  \bibfield  {author} {\bibinfo {author} {\bibfnamefont {R.}~\bibnamefont
  {Smola{\'n}czuk}},\ }\href@noop {} {\bibfield  {journal} {\bibinfo  {journal}
  {Phys. Rev. C}\ }\textbf {\bibinfo {volume} {56}},\ \bibinfo {pages} {812}
  (\bibinfo {year} {1997})}\BibitemShut {NoStop}%
\bibitem [{\citenamefont {Gherghescu}\ \emph {et~al.}(1999)\citenamefont
  {Gherghescu}, \citenamefont {Skalski}, \citenamefont {Patyk},\ and\
  \citenamefont {Sobiczewski}}]{Ghe99}%
  \BibitemOpen
  \bibfield  {author} {\bibinfo {author} {\bibfnamefont {R.}~\bibnamefont
  {Gherghescu}}, \bibinfo {author} {\bibfnamefont {J.}~\bibnamefont {Skalski}},
  \bibinfo {author} {\bibfnamefont {Z.}~\bibnamefont {Patyk}}, \ and\ \bibinfo
  {author} {\bibfnamefont {A.}~\bibnamefont {Sobiczewski}},\ }\href@noop {}
  {\bibfield  {journal} {\bibinfo  {journal} {Nucl. Phys. A}\ }\textbf
  {\bibinfo {volume} {651}},\ \bibinfo {pages} {237} (\bibinfo {year}
  {1999})}\BibitemShut {NoStop}%
\bibitem [{\citenamefont {Erler}\ \emph {et~al.}(2012)\citenamefont {Erler},
  \citenamefont {Langanke}, \citenamefont {Loens}, \citenamefont
  {Martinez-Pinedo},\ and\ \citenamefont {Reinhard}}]{Erl12}%
  \BibitemOpen
  \bibfield  {author} {\bibinfo {author} {\bibfnamefont {J.}~\bibnamefont
  {Erler}}, \bibinfo {author} {\bibfnamefont {K.}~\bibnamefont {Langanke}},
  \bibinfo {author} {\bibfnamefont {H.}~\bibnamefont {Loens}}, \bibinfo
  {author} {\bibfnamefont {G.}~\bibnamefont {Martinez-Pinedo}}, \ and\ \bibinfo
  {author} {\bibfnamefont {P.-G.}\ \bibnamefont {Reinhard}},\ }\href@noop {}
  {\bibfield  {journal} {\bibinfo  {journal} {Phys. Rev. C}\ }\textbf {\bibinfo
  {volume} {85}},\ \bibinfo {pages} {025802} (\bibinfo {year}
  {2012})}\BibitemShut {NoStop}%
\bibitem [{\citenamefont {Schindzielorz}\ \emph {et~al.}(2009)\citenamefont
  {Schindzielorz}, \citenamefont {Erler}, \citenamefont {Kl{\"u}pfel},\ and\
  \citenamefont {Reinhard}}]{Schin09}%
  \BibitemOpen
  \bibfield  {author} {\bibinfo {author} {\bibfnamefont {N.}~\bibnamefont
  {Schindzielorz}}, \bibinfo {author} {\bibfnamefont {J.}~\bibnamefont
  {Erler}}, \bibinfo {author} {\bibfnamefont {P.}~\bibnamefont {Kl{\"u}pfel}},
  \ and\ \bibinfo {author} {\bibfnamefont {P.-G.}\ \bibnamefont {Reinhard}},\
  }\href@noop {} {\bibfield  {journal} {\bibinfo  {journal} {Int. J. Mod. Phys.
  E}\ }\textbf {\bibinfo {volume} {18}},\ \bibinfo {pages} {773} (\bibinfo
  {year} {2009})}\BibitemShut {NoStop}%
\bibitem [{\citenamefont {Warda}\ and\ \citenamefont {Egido}(2012)}]{Warda12}%
  \BibitemOpen
  \bibfield  {author} {\bibinfo {author} {\bibfnamefont {M.}~\bibnamefont
  {Warda}}\ and\ \bibinfo {author} {\bibfnamefont {J.}~\bibnamefont {Egido}},\
  }\href@noop {} {\  (\bibinfo {year} {2012})},\ \Eprint
  {http://arxiv.org/abs/nucl-th/1204.5867} {arXiv:nucl-th/1204.5867}
  \BibitemShut {NoStop}%
\bibitem [{\citenamefont {Krappe}\ and\ \citenamefont
  {Pomorski}(2012)}]{Krappe-Pomorski}%
  \BibitemOpen
  \bibfield  {author} {\bibinfo {author} {\bibfnamefont {H.~J.}\ \bibnamefont
  {Krappe}}\ and\ \bibinfo {author} {\bibfnamefont {K.}~\bibnamefont
  {Pomorski}},\ }\href@noop {} {\emph {\bibinfo {title} {Theory of Nuclear
  Fission: A Textbook}}},\ \bibinfo {series} {Lecture Notes in Physics}, Vol.\
  \bibinfo {volume} {838}\ (\bibinfo  {publisher} {Springer},\ \bibinfo {year}
  {2012})\BibitemShut {NoStop}%
\bibitem [{\citenamefont {Bender}\ \emph {et~al.}(2003)\citenamefont {Bender},
  \citenamefont {Heenen},\ and\ \citenamefont {Reinhard}}]{Ben03}%
  \BibitemOpen
  \bibfield  {author} {\bibinfo {author} {\bibfnamefont {M.}~\bibnamefont
  {Bender}}, \bibinfo {author} {\bibfnamefont {P.-H.}\ \bibnamefont {Heenen}},
  \ and\ \bibinfo {author} {\bibfnamefont {P.-G.}\ \bibnamefont {Reinhard}},\
  }\href@noop {} {\bibfield  {journal} {\bibinfo  {journal} {Rev. Mod. Phys.}\
  }\textbf {\bibinfo {volume} {75}},\ \bibinfo {pages} {121} (\bibinfo {year}
  {2003})}\BibitemShut {NoStop}%
\bibitem [{\citenamefont {Goutte}\ \emph {et~al.}(2005)\citenamefont {Goutte},
  \citenamefont {Berger}, \citenamefont {Casoli},\ and\ \citenamefont
  {Gogny}}]{Gou05}%
  \BibitemOpen
  \bibfield  {author} {\bibinfo {author} {\bibfnamefont {H.}~\bibnamefont
  {Goutte}}, \bibinfo {author} {\bibfnamefont {J.}~\bibnamefont {Berger}},
  \bibinfo {author} {\bibfnamefont {P.}~\bibnamefont {Casoli}}, \ and\ \bibinfo
  {author} {\bibfnamefont {D.}~\bibnamefont {Gogny}},\ }\href@noop {}
  {\bibfield  {journal} {\bibinfo  {journal} {Phys. Rev. C}\ }\textbf {\bibinfo
  {volume} {71}},\ \bibinfo {pages} {024316} (\bibinfo {year}
  {2005})}\BibitemShut {NoStop}%
\bibitem [{\citenamefont {Baranger}\ and\ \citenamefont
  {V\'en\'eroni}(1978)}]{Bar78}%
  \BibitemOpen
  \bibfield  {author} {\bibinfo {author} {\bibfnamefont {M.}~\bibnamefont
  {Baranger}}\ and\ \bibinfo {author} {\bibfnamefont {M.}~\bibnamefont
  {V\'en\'eroni}},\ }\href@noop {} {\bibfield  {journal} {\bibinfo  {journal}
  {Ann. Phys.}\ }\textbf {\bibinfo {volume} {114}},\ \bibinfo {pages} {123}
  (\bibinfo {year} {1978})}\BibitemShut {NoStop}%
\bibitem [{\citenamefont {Dobaczewski}\ and\ \citenamefont
  {Skalski}(1981)}]{Dob81}%
  \BibitemOpen
  \bibfield  {author} {\bibinfo {author} {\bibfnamefont {J.}~\bibnamefont
  {Dobaczewski}}\ and\ \bibinfo {author} {\bibfnamefont {J.}~\bibnamefont
  {Skalski}},\ }\href@noop {} {\bibfield  {journal} {\bibinfo  {journal} {Nucl.
  Phys. A}\ }\textbf {\bibinfo {volume} {369}},\ \bibinfo {pages} {123}
  (\bibinfo {year} {1981})}\BibitemShut {NoStop}%
\bibitem [{\citenamefont {Gr{\"u}mmer}\ \emph {et~al.}(1982)\citenamefont
  {Gr{\"u}mmer}, \citenamefont {Goeke},\ and\ \citenamefont
  {Reinhard}}]{Gru82}%
  \BibitemOpen
  \bibfield  {author} {\bibinfo {author} {\bibfnamefont {F.}~\bibnamefont
  {Gr{\"u}mmer}}, \bibinfo {author} {\bibfnamefont {K.}~\bibnamefont {Goeke}},
  \ and\ \bibinfo {author} {\bibfnamefont {P.-G.}\ \bibnamefont {Reinhard}},\
  }\href@noop {} {\bibfield  {journal} {\bibinfo  {journal} {Lecture Notes in
  Physics}\ }\textbf {\bibinfo {volume} {171}},\ \bibinfo {pages} {323}
  (\bibinfo {year} {1982})}\BibitemShut {NoStop}%
\bibitem [{\citenamefont {Skalski}(2008)}]{Ska08}%
  \BibitemOpen
  \bibfield  {author} {\bibinfo {author} {\bibfnamefont {J.}~\bibnamefont
  {Skalski}},\ }\href@noop {} {\bibfield  {journal} {\bibinfo  {journal} {Phys.
  Rev. C}\ }\textbf {\bibinfo {volume} {77}},\ \bibinfo {pages} {064610}
  (\bibinfo {year} {2008})}\BibitemShut {NoStop}%
\bibitem [{\citenamefont {Yuldashbaeva}\ \emph {et~al.}(1999)\citenamefont
  {Yuldashbaeva}, \citenamefont {Libert}, \citenamefont {Quentin},\ and\
  \citenamefont {Girod}}]{Yul99all}%
  \BibitemOpen
  \bibfield  {author} {\bibinfo {author} {\bibfnamefont {E.}~\bibnamefont
  {Yuldashbaeva}}, \bibinfo {author} {\bibfnamefont {J.}~\bibnamefont
  {Libert}}, \bibinfo {author} {\bibfnamefont {P.}~\bibnamefont {Quentin}}, \
  and\ \bibinfo {author} {\bibfnamefont {M.}~\bibnamefont {Girod}},\
  }\href@noop {} {\bibfield  {journal} {\bibinfo  {journal} {Phys. Lett. B}\
  }\textbf {\bibinfo {volume} {461}},\ \bibinfo {pages} {1} (\bibinfo {year}
  {1999})}\BibitemShut {NoStop}%
\bibitem [{\citenamefont {Baran}\ \emph {et~al.}(2011)\citenamefont {Baran},
  \citenamefont {Sheikh}, \citenamefont {Dobaczewski}, \citenamefont
  {Nazarewicz},\ and\ \citenamefont {Staszczak}}]{Baran11call}%
  \BibitemOpen
  \bibfield  {author} {\bibinfo {author} {\bibfnamefont {A.}~\bibnamefont
  {Baran}}, \bibinfo {author} {\bibfnamefont {J.~A.}\ \bibnamefont {Sheikh}},
  \bibinfo {author} {\bibfnamefont {J.}~\bibnamefont {Dobaczewski}}, \bibinfo
  {author} {\bibfnamefont {W.}~\bibnamefont {Nazarewicz}}, \ and\ \bibinfo
  {author} {\bibfnamefont {A.}~\bibnamefont {Staszczak}},\ }\href@noop {}
  {\bibfield  {journal} {\bibinfo  {journal} {Phys. Rev. C}\ }\textbf {\bibinfo
  {volume} {84}},\ \bibinfo {pages} {054321} (\bibinfo {year}
  {2011})}\BibitemShut {NoStop}%
\bibitem [{\citenamefont {Staszczak}\ \emph
  {et~al.}(2011{\natexlab{b}})\citenamefont {Staszczak}, \citenamefont
  {Baran},\ and\ \citenamefont {Nazarewicz}}]{Sta11}%
  \BibitemOpen
  \bibfield  {author} {\bibinfo {author} {\bibfnamefont {A.}~\bibnamefont
  {Staszczak}}, \bibinfo {author} {\bibfnamefont {A.}~\bibnamefont {Baran}}, \
  and\ \bibinfo {author} {\bibfnamefont {W.}~\bibnamefont {Nazarewicz}},\
  }\href@noop {} {\bibfield  {journal} {\bibinfo  {journal} {Int. J. Mod. Phys.
  E}\ }\textbf {\bibinfo {volume} {20}},\ \bibinfo {pages} {552} (\bibinfo
  {year} {2011}{\natexlab{b}})}\BibitemShut {NoStop}%
\bibitem [{\citenamefont {Schunck}\ \emph {et~al.}(2012)\citenamefont
  {Schunck}, \citenamefont {Dobaczewski}, \citenamefont {McDonnell},
  \citenamefont {Satu{\l}a}, \citenamefont {Sheikh}, \citenamefont {Staszczak},
  \citenamefont {Stoitsov},\ and\ \citenamefont {Toivanen}}]{hfodd7all}%
  \BibitemOpen
  \bibfield  {author} {\bibinfo {author} {\bibfnamefont {N.}~\bibnamefont
  {Schunck}}, \bibinfo {author} {\bibfnamefont {J.}~\bibnamefont
  {Dobaczewski}}, \bibinfo {author} {\bibfnamefont {J.}~\bibnamefont
  {McDonnell}}, \bibinfo {author} {\bibfnamefont {W.}~\bibnamefont
  {Satu{\l}a}}, \bibinfo {author} {\bibfnamefont {J.~A.}\ \bibnamefont
  {Sheikh}}, \bibinfo {author} {\bibfnamefont {A.}~\bibnamefont {Staszczak}},
  \bibinfo {author} {\bibfnamefont {M.}~\bibnamefont {Stoitsov}}, \ and\
  \bibinfo {author} {\bibfnamefont {P.}~\bibnamefont {Toivanen}},\ }\href@noop
  {} {\bibfield  {journal} {\bibinfo  {journal} {Comput. Phys. Commun.}\
  }\textbf {\bibinfo {volume} {183}},\ \bibinfo {pages} {166} (\bibinfo {year}
  {2012})}\BibitemShut {NoStop}%
\bibitem [{\citenamefont {Bartel}\ \emph {et~al.}(1982)\citenamefont {Bartel},
  \citenamefont {Quentin}, \citenamefont {Brack}, \citenamefont {Guet},\ and\
  \citenamefont {H{\aa}kansson}}]{Bart82all}%
  \BibitemOpen
  \bibfield  {author} {\bibinfo {author} {\bibfnamefont {J.}~\bibnamefont
  {Bartel}}, \bibinfo {author} {\bibfnamefont {P.}~\bibnamefont {Quentin}},
  \bibinfo {author} {\bibfnamefont {M.}~\bibnamefont {Brack}}, \bibinfo
  {author} {\bibfnamefont {C.}~\bibnamefont {Guet}}, \ and\ \bibinfo {author}
  {\bibfnamefont {H.~B.}\ \bibnamefont {H{\aa}kansson}},\ }\href@noop {}
  {\bibfield  {journal} {\bibinfo  {journal} {Nucl. Phys. A}\ }\textbf
  {\bibinfo {volume} {386}},\ \bibinfo {pages} {79} (\bibinfo {year}
  {1982})}\BibitemShut {NoStop}%
\bibitem [{\citenamefont {Kortelainen}\ \emph {et~al.}(2012)\citenamefont
  {Kortelainen}, \citenamefont {McDonnell}, \citenamefont {Nazarewicz},
  \citenamefont {Reinhard}, \citenamefont {Sarich}, \citenamefont {Schunck},
  \citenamefont {Stoitsov},\ and\ \citenamefont {Wild}}]{Kor12}%
  \BibitemOpen
  \bibfield  {author} {\bibinfo {author} {\bibfnamefont {M.}~\bibnamefont
  {Kortelainen}}, \bibinfo {author} {\bibfnamefont {J.}~\bibnamefont
  {McDonnell}}, \bibinfo {author} {\bibfnamefont {W.}~\bibnamefont
  {Nazarewicz}}, \bibinfo {author} {\bibfnamefont {P.-G.}\ \bibnamefont
  {Reinhard}}, \bibinfo {author} {\bibfnamefont {J.}~\bibnamefont {Sarich}},
  \bibinfo {author} {\bibfnamefont {N.}~\bibnamefont {Schunck}}, \bibinfo
  {author} {\bibfnamefont {M.~V.}\ \bibnamefont {Stoitsov}}, \ and\ \bibinfo
  {author} {\bibfnamefont {S.~M.}\ \bibnamefont {Wild}},\ }\href@noop {}
  {\bibfield  {journal} {\bibinfo  {journal} {Phys. Rev. C}\ }\textbf {\bibinfo
  {volume} {85}},\ \bibinfo {pages} {024304} (\bibinfo {year}
  {2012})}\BibitemShut {NoStop}%
\bibitem [{\citenamefont {McDonnell}(2012)}]{McDonn12}%
  \BibitemOpen
  \bibfield  {author} {\bibinfo {author} {\bibfnamefont {J.}~\bibnamefont
  {McDonnell}},\ }\href@noop {} {\bibfield  {journal} {\bibinfo  {journal}
  {Ph.D. Thesis, to be published}\ } (\bibinfo {year} {2012})}\BibitemShut
  {NoStop}%
\bibitem [{\citenamefont {Dobaczewski}\ \emph {et~al.}(2002)\citenamefont
  {Dobaczewski}, \citenamefont {Nazarewicz},\ and\ \citenamefont
  {Stoitsov}}]{Dob02}%
  \BibitemOpen
  \bibfield  {author} {\bibinfo {author} {\bibfnamefont {J.}~\bibnamefont
  {Dobaczewski}}, \bibinfo {author} {\bibfnamefont {W.}~\bibnamefont
  {Nazarewicz}}, \ and\ \bibinfo {author} {\bibfnamefont {M.}~\bibnamefont
  {Stoitsov}},\ }\href@noop {} {\bibfield  {journal} {\bibinfo  {journal} {Eur.
  Phys. J. A}\ }\textbf {\bibinfo {volume} {15}},\ \bibinfo {pages} {21}
  (\bibinfo {year} {2002})}\BibitemShut {NoStop}%
\bibitem [{\citenamefont {Dobaczewski}\ \emph {et~al.}(1984)\citenamefont
  {Dobaczewski}, \citenamefont {Flocard},\ and\ \citenamefont
  {Treiner}}]{Dob84}%
  \BibitemOpen
  \bibfield  {author} {\bibinfo {author} {\bibfnamefont {J.}~\bibnamefont
  {Dobaczewski}}, \bibinfo {author} {\bibfnamefont {H.}~\bibnamefont
  {Flocard}}, \ and\ \bibinfo {author} {\bibfnamefont {J.}~\bibnamefont
  {Treiner}},\ }\href@noop {} {\bibfield  {journal} {\bibinfo  {journal} {Nucl.
  Phys. A}\ }\textbf {\bibinfo {volume} {422}} (\bibinfo {year}
  {1984})}\BibitemShut {NoStop}%
\bibitem [{\citenamefont {Dobaczewski}\ \emph {et~al.}(1996)\citenamefont
  {Dobaczewski}, \citenamefont {Nazarewicz}, \citenamefont {Werner},
  \citenamefont {Berger}, \citenamefont {Chinn},\ and\ \citenamefont
  {Decharg\'e}}]{Dob96all}%
  \BibitemOpen
  \bibfield  {author} {\bibinfo {author} {\bibfnamefont {J.}~\bibnamefont
  {Dobaczewski}}, \bibinfo {author} {\bibfnamefont {W.}~\bibnamefont
  {Nazarewicz}}, \bibinfo {author} {\bibfnamefont {T.~R.}\ \bibnamefont
  {Werner}}, \bibinfo {author} {\bibfnamefont {J.-F.}\ \bibnamefont {Berger}},
  \bibinfo {author} {\bibfnamefont {C.}~\bibnamefont {Chinn}}, \ and\ \bibinfo
  {author} {\bibfnamefont {J.}~\bibnamefont {Decharg\'e}},\ }\href@noop {}
  {\bibfield  {journal} {\bibinfo  {journal} {Phys. Rev. C}\ }\textbf {\bibinfo
  {volume} {53}},\ \bibinfo {pages} {2809} (\bibinfo {year}
  {1996})}\BibitemShut {NoStop}%
\bibitem [{\citenamefont {Borycki}\ \emph {et~al.}(2006)\citenamefont
  {Borycki}, \citenamefont {Dobaczewski}, \citenamefont {Nazarewicz},\ and\
  \citenamefont {Stoitsov}}]{Bor06awall}%
  \BibitemOpen
  \bibfield  {author} {\bibinfo {author} {\bibfnamefont {P.~J.}\ \bibnamefont
  {Borycki}}, \bibinfo {author} {\bibfnamefont {J.}~\bibnamefont
  {Dobaczewski}}, \bibinfo {author} {\bibfnamefont {W.}~\bibnamefont
  {Nazarewicz}}, \ and\ \bibinfo {author} {\bibfnamefont {M.}~\bibnamefont
  {Stoitsov}},\ }\href@noop {} {\bibfield  {journal} {\bibinfo  {journal}
  {Phys. Rev. C}\ }\textbf {\bibinfo {volume} {73}},\ \bibinfo {pages} {044319}
  (\bibinfo {year} {2006})}\BibitemShut {NoStop}%
\bibitem [{\citenamefont {Pei}\ \emph {et~al.}(2011)\citenamefont {Pei},
  \citenamefont {Kruppa},\ and\ \citenamefont {Nazarewicz}}]{Pei11}%
  \BibitemOpen
  \bibfield  {author} {\bibinfo {author} {\bibfnamefont {J.}~\bibnamefont
  {Pei}}, \bibinfo {author} {\bibfnamefont {A.}~\bibnamefont {Kruppa}}, \ and\
  \bibinfo {author} {\bibfnamefont {W.}~\bibnamefont {Nazarewicz}},\
  }\href@noop {} {\bibfield  {journal} {\bibinfo  {journal} {Phys. Rev. C}\
  }\textbf {\bibinfo {volume} {84}},\ \bibinfo {pages} {024311} (\bibinfo
  {year} {2011})}\BibitemShut {NoStop}%
\bibitem [{\citenamefont {Staszczak}\ \emph {et~al.}(2005)\citenamefont
  {Staszczak}, \citenamefont {Dobaczewski},\ and\ \citenamefont
  {Nazarewicz}}]{Sta05a}%
  \BibitemOpen
  \bibfield  {author} {\bibinfo {author} {\bibfnamefont {A.}~\bibnamefont
  {Staszczak}}, \bibinfo {author} {\bibfnamefont {J.}~\bibnamefont
  {Dobaczewski}}, \ and\ \bibinfo {author} {\bibfnamefont {W.}~\bibnamefont
  {Nazarewicz}},\ }\href@noop {} {\bibfield  {journal} {\bibinfo  {journal}
  {Int. J. Mod. Phys. E}\ }\textbf {\bibinfo {volume} {14}},\ \bibinfo {pages}
  {395} (\bibinfo {year} {2005})}\BibitemShut {NoStop}%
\bibitem [{\citenamefont {Staszczak}\ \emph {et~al.}(1989)\citenamefont
  {Staszczak}, \citenamefont {Pi{\l}at},\ and\ \citenamefont
  {Pomorski}}]{Staszczak89}%
  \BibitemOpen
  \bibfield  {author} {\bibinfo {author} {\bibfnamefont {A.}~\bibnamefont
  {Staszczak}}, \bibinfo {author} {\bibfnamefont {S.}~\bibnamefont {Pi{\l}at}},
  \ and\ \bibinfo {author} {\bibfnamefont {K.}~\bibnamefont {Pomorski}},\
  }\href@noop {} {\bibfield  {journal} {\bibinfo  {journal} {Nucl. Phys. A}\
  }\textbf {\bibinfo {volume} {504}},\ \bibinfo {pages} {589} (\bibinfo {year}
  {1989})}\BibitemShut {NoStop}%
\bibitem [{\citenamefont {Baran}\ \emph {et~al.}(2007)\citenamefont {Baran},
  \citenamefont {Staszczak}, \citenamefont {Dobaczewski},\ and\ \citenamefont
  {Nazarewicz}}]{Baran07call}%
  \BibitemOpen
  \bibfield  {author} {\bibinfo {author} {\bibfnamefont {A.}~\bibnamefont
  {Baran}}, \bibinfo {author} {\bibfnamefont {A.}~\bibnamefont {Staszczak}},
  \bibinfo {author} {\bibfnamefont {J.}~\bibnamefont {Dobaczewski}}, \ and\
  \bibinfo {author} {\bibfnamefont {W.}~\bibnamefont {Nazarewicz}},\
  }\href@noop {} {\bibfield  {journal} {\bibinfo  {journal} {Int. J. Mod. Phys.
  E}\ }\textbf {\bibinfo {volume} {16}},\ \bibinfo {pages} {443} (\bibinfo
  {year} {2007})}\BibitemShut {NoStop}%
\bibitem [{\citenamefont {Leboeuf}\ and\ \citenamefont
  {Sharma}(1973{\natexlab{a}})}]{Leboeuf73a}%
  \BibitemOpen
  \bibfield  {author} {\bibinfo {author} {\bibfnamefont {J.~N.}\ \bibnamefont
  {Leboeuf}}\ and\ \bibinfo {author} {\bibfnamefont {R.~C.}\ \bibnamefont
  {Sharma}},\ }\href@noop {} {\bibfield  {journal} {\bibinfo  {journal} {Can.
  J. Phys.}\ }\textbf {\bibinfo {volume} {51}},\ \bibinfo {pages} {446}
  (\bibinfo {year} {1973}{\natexlab{a}})}\BibitemShut {NoStop}%
\bibitem [{\citenamefont {Leboeuf}\ and\ \citenamefont
  {Sharma}(1973{\natexlab{b}})}]{Leboeuf73b}%
  \BibitemOpen
  \bibfield  {author} {\bibinfo {author} {\bibfnamefont {J.~N.}\ \bibnamefont
  {Leboeuf}}\ and\ \bibinfo {author} {\bibfnamefont {R.~C.}\ \bibnamefont
  {Sharma}},\ }\href@noop {} {\bibfield  {journal} {\bibinfo  {journal} {Can.
  J. Phys.}\ }\textbf {\bibinfo {volume} {51}},\ \bibinfo {pages} {1148}
  (\bibinfo {year} {1973}{\natexlab{b}})}\BibitemShut {NoStop}%
\bibitem [{\citenamefont {Leboeuf}\ and\ \citenamefont
  {Sharma}(1973{\natexlab{c}})}]{Leboeuf73c}%
  \BibitemOpen
  \bibfield  {author} {\bibinfo {author} {\bibfnamefont {J.~N.}\ \bibnamefont
  {Leboeuf}}\ and\ \bibinfo {author} {\bibfnamefont {R.~C.}\ \bibnamefont
  {Sharma}},\ }\href@noop {} {\bibfield  {journal} {\bibinfo  {journal} {Can.
  J. Phys.}\ }\textbf {\bibinfo {volume} {51}},\ \bibinfo {pages} {2023}
  (\bibinfo {year} {1973}{\natexlab{c}})}\BibitemShut {NoStop}%
\bibitem [{\citenamefont {Holden}\ and\ \citenamefont {Hoffman}(2000)}]{Hol00}%
  \BibitemOpen
  \bibfield  {author} {\bibinfo {author} {\bibfnamefont {N.}~\bibnamefont
  {Holden}}\ and\ \bibinfo {author} {\bibfnamefont {D.}~\bibnamefont
  {Hoffman}},\ }\href@noop {} {\bibfield  {journal} {\bibinfo  {journal} {Pure
  Appl. Chem.}\ }\textbf {\bibinfo {volume} {72}},\ \bibinfo {pages} {1525}
  (\bibinfo {year} {2000})}\BibitemShut {NoStop}%
\bibitem [{\citenamefont {Khuyagbaatar}\ \emph {et~al.}(2008)\citenamefont
  {Khuyagbaatar} \emph {et~al.}}]{Khu08}%
  \BibitemOpen
  \bibfield  {author} {\bibinfo {author} {\bibfnamefont {J.}~\bibnamefont
  {Khuyagbaatar}} \emph {et~al.},\ }\href@noop {} {\bibfield  {journal}
  {\bibinfo  {journal} {Eur. Phys. J. A}\ }\textbf {\bibinfo {volume} {37}},\
  \bibinfo {pages} {177} (\bibinfo {year} {2008})}\BibitemShut {NoStop}%
\bibitem [{\citenamefont {Jachimowicz}\ \emph {et~al.}(2011)\citenamefont
  {Jachimowicz}, \citenamefont {Kowal},\ and\ \citenamefont
  {Skalski}}]{Jach11}%
  \BibitemOpen
  \bibfield  {author} {\bibinfo {author} {\bibfnamefont {P.}~\bibnamefont
  {Jachimowicz}}, \bibinfo {author} {\bibfnamefont {M.}~\bibnamefont {Kowal}},
  \ and\ \bibinfo {author} {\bibfnamefont {J.}~\bibnamefont {Skalski}},\
  }\href@noop {} {\bibfield  {journal} {\bibinfo  {journal} {Phys. Rev. C}\
  }\textbf {\bibinfo {volume} {83}},\ \bibinfo {pages} {054302} (\bibinfo
  {year} {2011})}\BibitemShut {NoStop}%
\bibitem [{\citenamefont {M{\"o}ller}\ \emph {et~al.}(2012)\citenamefont
  {M{\"o}ller}, \citenamefont {Sierk}, \citenamefont {Bentsson}, \citenamefont
  {Sagawa},\ and\ \citenamefont {Ichikawa}}]{Mol12t}%
  \BibitemOpen
  \bibfield  {author} {\bibinfo {author} {\bibfnamefont {P.}~\bibnamefont
  {M{\"o}ller}}, \bibinfo {author} {\bibfnamefont {A.}~\bibnamefont {Sierk}},
  \bibinfo {author} {\bibfnamefont {R.}~\bibnamefont {Bentsson}}, \bibinfo
  {author} {\bibfnamefont {H.}~\bibnamefont {Sagawa}}, \ and\ \bibinfo {author}
  {\bibfnamefont {T.}~\bibnamefont {Ichikawa}},\ }\href@noop {} {\bibfield
  {journal} {\bibinfo  {journal} {At. Data Nucl. data Tables}\ }\textbf
  {\bibinfo {volume} {98}},\ \bibinfo {pages} {149} (\bibinfo {year}
  {2012})}\BibitemShut {NoStop}%
\bibitem [{\citenamefont {{{Viola, Jr.}}}\ and\ \citenamefont
  {Seaborg}(1966)}]{Viola66}%
  \BibitemOpen
  \bibfield  {author} {\bibinfo {author} {\bibfnamefont {V.~E.}\ \bibnamefont
  {{{Viola, Jr.}}}}\ and\ \bibinfo {author} {\bibfnamefont {G.~T.}\
  \bibnamefont {Seaborg}},\ }\href@noop {} {\bibfield  {journal} {\bibinfo
  {journal} {J. Inorg. Nucl. Chem.}\ }\textbf {\bibinfo {volume} {28}},\
  \bibinfo {pages} {741} (\bibinfo {year} {1966})}\BibitemShut {NoStop}%
\bibitem [{\citenamefont {Parkhomenko}\ and\ \citenamefont
  {Sobiczewski}(2005)}]{Sobi05}%
  \BibitemOpen
  \bibfield  {author} {\bibinfo {author} {\bibfnamefont {A.}~\bibnamefont
  {Parkhomenko}}\ and\ \bibinfo {author} {\bibfnamefont {A.}~\bibnamefont
  {Sobiczewski}},\ }\href@noop {} {\bibfield  {journal} {\bibinfo  {journal}
  {Acta Phys. Pol. B}\ }\textbf {\bibinfo {volume} {36}},\ \bibinfo {pages}
  {3095} (\bibinfo {year} {2005})}\BibitemShut {NoStop}%
\end{thebibliography}
%

\providecommand{\noopsort}[1]{}\providecommand{\singleletter}[1]{#1}%

\end{document}